\documentstyle[prl,aps,preprint]{revtex}

\begin{document}

\bigskip\ 

\bigskip\ 

\begin{center}
{\bf p-BRANE ACTION FROM GRAVI-DILATON }

{\bf EFFECTIVE ACTION}

{\bf \ }

{\bf \ }

\smallskip\ 

J. A. Nieto\footnote{%
nieto@uas.uasnet.mx} and C. M Yee

\smallskip

{\it Facultad de Ciencias F\'{\i}sico-Matem\'{a}ticas de la Universidad
Aut\'{o}noma}

{\it de Sinaloa, C.P. 80010, Culiac\'{a}n Sinaloa, M\'{e}xico}

\bigskip\ 

\bigskip\ 

{\bf Abstract}
\end{center}

Using a special ansatz for the metric, by straightforward computation we
prove that gravi-dilaton effective action in higher dimensions is reduced to
the p-brane action. The dual symmetry of the generic type $%
a\longleftrightarrow \frac 1a$ is an important symmetry of the reduced
action.

\smallskip\ 

Pacs numbers: 11.10.Kk, 04.50.+h, 12.60.-i

July/2000

\newpage\ 

In this brief note, by a straightforward computation we prove the surprising
result that using a particular ansatz for the metric the gravi-dilaton
action in $d+1$ dimensions is reduced to the $p-$brane action. This result
may be of special interest in the Randall-Sundrum scenario [1-2], string
theory [3] and M-theory [4].

Our starting point is the graviton-dilaton effective action with
cosmological constant:

\begin{equation}
S=-%
{\textstyle{1 \over 16\pi G_{d+1}}}%
\int d^{d+1}y\sqrt{-g}e^{-\phi }(R+(\nabla \phi )^2+2\Lambda ),  \label{1}
\end{equation}
where $G_{d+1\text{ }}$is the Newton constant in $d+1$ dimensions, $\phi
=\phi (y^\alpha )$ is the dilaton field and $R$ is the Ricci scalar obtained
from the Riemann tensor

\begin{equation}
R_{\nu \alpha \beta }^\mu =\Gamma _{\nu \beta ,\alpha }^\mu -\Gamma _{\nu
\alpha ,\beta }^\mu +\Gamma _{\sigma \alpha }^\mu \Gamma _{\nu \beta
}^\sigma -\Gamma _{\sigma \beta }^\mu \Gamma _{\nu \alpha }^\sigma  \label{2}
\end{equation}
and the metric tensor $g_{\alpha \beta }$, with $\alpha ,\beta =0,1...,d$.
Here, $\Gamma _{\alpha \beta }^\mu $ is the Christoffel symbol:

\begin{equation}
\Gamma _{\alpha \beta }^\mu =\frac 12g^{\mu \nu }(g_{\nu \alpha ,\beta
}+g_{\nu \beta ,\alpha }-g_{\alpha \beta ,\nu }).  \label{3}
\end{equation}

Consider the ansatz

\begin{equation}
\begin{array}{ccc}
g_{AB} & = & \tilde{g}_{AB}(y^{C}), \\ 
&  &  \\ 
g_{ij} & = & a_{k}(y^{C})a_{l}(y^{C})\eta _{ij}^{kl}, \\ 
&  &  \\ 
g_{Ai} & = & 0.
\end{array}
\label{4}
\end{equation}
Here, the indices $A,B,...etc.$ run from $0$ to $p,$ the indices $%
i,j,...etc. $ run from $p+1$ to $d$ and the only non-vanishing terms of $%
\eta _{ij}^{kl}$ are

\begin{equation}
\eta _{ij}^{kl}=1\text{, when }k=l=i=j.  \label{5}
\end{equation}

Assume that

\begin{equation}
\phi =\phi (y^C).  \label{6}
\end{equation}
From (3), (4) and (6) we find that the only non-vanishing Christoffel
symbols are

\begin{equation}
\begin{array}{ccc}
\Gamma _{ij}^A & = & -a_k\partial ^Aa_l\eta _{ij}^{kl}, \\ 
&  &  \\ 
\Gamma _{jA}^i & = & a^k\partial _Aa_l\eta _{kj}^{li}, \\ 
&  &  \\ 
\Gamma _{BC}^A & = & \tilde{\Gamma}_{BC}^A,
\end{array}
\label{7}
\end{equation}
where $\tilde{\Gamma}_{BC}^A$ is the Christoffel symbol associated to $%
\tilde{g}_{AB}.$ Here, $a^i=a_i^{-1},$ so we can take $g^{ij}=a^ka^l\eta
_{kl}^{ij}.$

From (2), (4) and (7) we discover that the only non-vanishing components of
the Riemann tensor are

\begin{equation}
\begin{array}{ccc}
R_{BCD}^A & = & \tilde{R}_{BCD}^A, \\ 
&  &  \\ 
R_{iBj}^A & = & -a_kD_B\partial ^Aa_l\eta _{ij}^{kl}, \\ 
&  &  \\ 
R_{AjB}^i & = & -a^kD_B\partial _Aa_l\eta _{kj}^{li}, \\ 
&  &  \\ 
R_{jkl}^i & = & -a^m\partial ^Aa_na_r\partial _Aa_s(\eta _{mk}^{ni}\eta
_{jl}^{rs}-\eta _{ml}^{ni}\eta _{jk}^{rs}).
\end{array}
\label{8}
\end{equation}
where $D_A$ denotes a covariant derivative in terms of $\tilde{\Gamma}%
_{BC}^A $. From (8) we find that the non-vanishing components of the Ricci
tensor $R_{\mu \nu }\equiv R_{\mu \alpha \nu }^\alpha $ are

\begin{equation}
\begin{array}{ccc}
R_{AB} & = & \tilde{R}_{AB}-a^iD_B\partial _Aa_i, \\ 
&  &  \\ 
R_{ij} & = & -(a_kD^A\partial _Aa_l+a^ma_k\partial _Aa_m\partial
^Aa_l-\partial _Aa_k\partial ^Aa_l)\eta _{ij}^{kl}.
\end{array}
\label{9}
\end{equation}
Thus, the Ricci scalar $R=g^{\mu \nu }R_{\mu \nu \text{ }}$is given by

\begin{equation}
\begin{array}{ccc}
R & = & -2a^iD^A\partial _Aa_i-a^ia^j\partial _Aa_i\partial ^Aa_j \\ 
&  &  \\ 
&  & +a^ia^j\partial _Aa_k\partial ^Aa_l\eta _{ij}^{kl}+\tilde{R}.
\end{array}
\label{10}
\end{equation}

Therefore, the action (1) becomes

\begin{equation}
\begin{array}{ccc}
S & = & -\frac 1{16\pi G_{p+1}}\int d^{p+1}y\sqrt{-\tilde{g}}\Pi a_se^{-\phi
}\{-2a^iD^A\partial _Aa_i-a^ia^j\partial ^Aa_i\partial _Aa_j \\ 
&  &  \\ 
&  & +a^ia^j\partial ^Aa_k\partial _Aa_l\eta _{ij}^{kl}+\partial ^A\phi
\partial _A\phi +\tilde{R}+2\Lambda \},
\end{array}
\label{11}
\end{equation}
where $G_{p+1\text{ }}$is the Newton constant in $p+1$ dimensions. The
relation between $G_{p+1\text{ }}$and $G_{d+1\text{ }}$is

\begin{equation}
\frac 1{G_{p+1\text{ }}}=\frac{V_n}{G_{d+1\text{ }}},  \label{12}
\end{equation}
where $V_n$ is a volume element in $n=d-p$ dimensions. This action can be
rewritten as

\begin{equation}
\begin{array}{c}
S=-\frac{1}{16\pi G_{p+1}}\int d^{p+1}y\sqrt{-\tilde{g}}D^{A}(-2\Pi
a_{s}e^{-\phi }a^{i}\partial _{A}a_{i}) \\ 
\\ 
-\frac{1}{16\pi G_{p+1}}\int d^{p+1}y\sqrt{-\tilde{g}}\Pi a_{s}e^{-\phi
}\{a^{i}a^{j}\partial ^{A}a_{i}\partial _{A}a_{j}-2\partial ^{A}\phi
a^{i}\partial _{A}a_{i} \\ 
\\ 
+\partial ^{A}\phi \partial _{A}\phi -a^{i}a^{j}\partial ^{A}a_{k}\partial
_{A}a_{l}\eta _{ij}^{kl}+2\Lambda \}-\frac{1}{16\pi G_{p+1}}\int d^{p+1}y%
\sqrt{-\tilde{g}}\Pi a_{s}e^{-\phi }\tilde{R}.
\end{array}
\label{13}
\end{equation}
Classically, since the first term in (13) is a total derivative we can be
drop it. Therefore, we have

\begin{equation}
\begin{array}{c}
S=-\frac 1{16\pi G_{p+1}}\int d^{p+1}y\sqrt{-\tilde{g}}\Pi a_se^{-\phi
}\{a^ia^j\partial ^Aa_i\partial _Aa_j-2\partial ^A\phi a^i\partial _Aa_i+ \\ 
\\ 
+\partial ^A\phi \partial _A\phi -a^ia^j\partial ^Aa_k\partial _Aa_l\eta
_{ij}^{kl}+2\Lambda \}-\frac 1{16\pi G_{p+1}}\int d^{p+1}y\sqrt{-\tilde{g}}%
\Pi a_se^{-\phi }\tilde{R}.
\end{array}
\label{14}
\end{equation}

Let us define $x^0$ as

\begin{equation}
\Pi a_se^{-\phi }=e^{-x^0}.  \label{15}
\end{equation}
We find

\begin{equation}
x^0=\phi -\sum \ln a_s  \label{16}
\end{equation}
and therefore,

\begin{equation}
\partial _Ax^0=\partial _A\phi -a^s\partial _Aa_s.  \label{17}
\end{equation}

Thus, the action (13) becomes

\begin{equation}
\begin{array}{c}
S=\frac 1{16\pi G_{p+1}}\int d^{p+1}y\sqrt{-\tilde{g}}e^{-x^0}\{-\partial
^Ax^0\partial _Ax^0+a^ia^j\partial ^Aa_k\partial _Aa_l\eta
_{ij}^{kl}-2\Lambda \} \\ 
\\ 
-\frac 1{16\pi G_{p+1}}\int d^{p+1}y\sqrt{-\tilde{g}}e^{-x^0}\tilde{R},
\end{array}
\label{18}
\end{equation}
\ It is not difficult to see that (18) is invariant under the duality
transformation

\[
a_i\longleftrightarrow \frac 1{a_i}. 
\]

Let us define the p-brane coupling ``constant'' $\Omega _p$ in the form

\begin{equation}
\frac{e^{-x^0}}{16\pi G_{p+1\text{ }}}=\frac 1{2\Omega _p}  \label{19}
\end{equation}
and the variables $x^i$ as

\begin{equation}
x^i\equiv \ln a_i.  \label{20}
\end{equation}
Using the expression (19) we find that (18) can be rewritten in the form

\begin{equation}
\begin{array}{ccc}
S & = & \frac 12\int \frac{d^{p+1}y}{\Omega _p}\sqrt{-\tilde{g}}\{-\partial
^Ax^0\partial _Ax^0+a^ia^j\partial ^Aa_k\partial _Aa_l\eta
_{ij}^{kl}-2\Lambda \} \\ 
&  &  \\ 
&  & -\frac 12\int \frac{d^{p+1}y}{\Omega _p}\sqrt{-\tilde{g}}\tilde{R}.
\end{array}
\label{21}
\end{equation}
If we now consider the definition (20) we find that (21) becomes

\begin{equation}
\begin{array}{ccc}
S & = & \frac 12\int \frac{d^{p+1}y}{\Omega _p}\sqrt{-\tilde{g}}\{-\partial
^Ax^0\partial _Ax^0+\partial ^Ax^i\partial _Ax^j\delta _{ij}-2\Lambda \}\}
\\ 
&  &  \\ 
&  & -\frac 12\int \frac{d^{p+1}y}{\Omega _p}\sqrt{-\tilde{g}}\tilde{R}.
\end{array}
\label{22}
\end{equation}
We easily note that (22) can be written as

\begin{equation}
\begin{array}{ccc}
S & = & \frac{1}{2}\int \frac{d^{p+1}y}{\Omega _{p}}\sqrt{-\tilde{g}}[\tilde{%
g}^{AB}\partial _{A}x^{\hat{\mu}}\partial _{B}x^{\hat{\nu}}\eta _{\hat{\mu}%
\hat{\nu}}-2\Lambda ] \\ 
&  &  \\ 
&  & -\frac{1}{2}\int \frac{d^{p+1}y}{\Omega _{p}}\sqrt{-\tilde{g}}\tilde{R},
\end{array}
\label{23}
\end{equation}
where $\eta _{\hat{\mu}\hat{\nu}}=diag(-1,1,...,1).$ Here the indices $\hat{%
\mu},\hat{\nu},...$etc run from $0$ to $n=d-p.$ By setting the cosmological
constant $\Lambda $ as

\begin{equation}
\Lambda =\frac{p-1}2,  \label{24}
\end{equation}
we recognize the action (23) as the p-brane action.

Let us make some final comments. We have shown explicitly that the p-brane
structure is contained in a higher dimensional effective gravi-dilaton
theory. In fact, our result is very general since applies to any higher
dimensional effective gravi-dilaton theory and any p-brane. The case of
0-brane (point particle) corresponds to a cosmological model and in fact
such a case has already been considered in the literature (see Ref. [5] and
references there in). The case of 1-brane, corresponding to strings, is of
great importance and deserve a special discussion.

From the point of view of the traditional string theory history our result
is clearly intriguing and surprising. Since for strings $p=1,$ from (24) we
see that $\Lambda =0$ and the action (23) is reduced to the well-known
Polyakov action. Let us assume that this reduced action is the bosonic
sector of the superstring action. We know that for superstrings, at the
quantum level, the Weyl invariance implies that $n+1=10.$ Therefore, in this
case the effective action (1) is defined in $d+1=11$ dimensions.
Consequently, (1) can be thought as the bosonic sector of eleven dimensional
supergravity and in this sense our result is in agreement with M-theory.

\bigskip\

\end{document}